%% file: ICSM20RestDep.tex
\documentclass[conference]{IEEEtran}
\usepackage{url}
\usepackage{balance}
\usepackage{booktabs}
\usepackage{listings}
\usepackage{algorithm}
\usepackage{algpseudocode}
\usepackage{caption}

\algrenewcommand\alglinenumber[1]{\tiny #1:}

\newboolean{showcomments}
\setboolean{showcomments}{true}
\ifthenelse{\boolean{showcomments}}
{ \newcommand{\mynote}[2]{\textcolor{red}{
			\fbox{\bfseries\sffamily\scriptsize#1}
			{\small$\blacktriangleright$\textsf{\emph{#2}}$\blacktriangleleft$}}}}
{ \newcommand{\mynote}[2]{}}

\newcommand{\yang}[1]{\mynote{Yang}{\textcolor{brown}{#1}}}

\newcommand{\squeezeup}{\vspace{-0.5cm}}

\usepackage{tcolorbox}

\newcommand{\rqone}{\textit{RQ1: To what extent do RESTFul APIs mention deprecation before introducing breaking changes?}}
\newcommand{\rqtwo}{\textit{RQ2: How do API providers deprecate RESTful APIs?}}
\newcommand{\rqthree}{\textit{RQ3: What deprecation-related information is provided for API consumers?}}
\newcommand{\rqfour}{\textit{RQ4: How API consumers are informed about the deprecation in RESTful APIs?}}

\ifCLASSOPTIONcompsoc
  \usepackage[nocompress]{cite}
\else
  \usepackage{cite}
\fi

\ifCLASSINFOpdf
\else
\fi
\begin{document}
%
\title{A First Look at the Deprecation of RESTful APIs: An Empirical Study
}

\author{\IEEEauthorblockN{Jerin Yasmin, Yuan Tian}
\IEEEauthorblockA{School of Computing\\
Queen's University\\
Kingston, Canada\\
Email: \{jerin.yasmin, y.tian\}@queensu.ca}
\and
\IEEEauthorblockN{Jinqiu Yang}
\IEEEauthorblockA{Department of Computer Science and Software Engineering\\
Concordia University\\
Montreal, Canada\\
Email: jinqiuy@encs.concordia.ca}
}

%


\pagestyle{plain}

\maketitle

\begin{abstract}
REpresentational State Transfer (REST) is considered as one standard software architectural style to build web APIs that can integrate software systems over the internet. However, while connecting systems, RESTful APIs might also break the dependent applications that rely on their services when they introduce breaking changes, e.g., an older version of the API is no longer supported. To warn developers promptly and thus prevent critical impact on downstream applications, a \textit{deprecated-removed} model should be followed, and deprecation-related information such as alternative approaches should also be listed. While API deprecation analysis as a theme is not new, most existing work focuses on non-web APIs, such as the ones provided by Java and Android.

To investigate RESTful API deprecation, we propose a framework called RADA (\underline{R}ESTful \underline{A}PI \underline{D}eprecation \underline{A}nalyzer). RADA is capable of automatically identifying deprecated API elements and analyzing impacted operations from an OpenAPI specification, a machine-readable profile for describing RESTful web service. We apply RADA on 2,224 OpenAPI specifications of 1,368 RESTful APIs collected from APIs.guru, the largest directory of OpenAPI specifications. Based on the data mined by RADA, we perform an empirical study to investigate how the deprecated-removed protocol is followed in RESTful APIs and characterize practices in RESTful API deprecation. The results of our study reveal several severe deprecation-related problems in existing RESTful APIs. Our implementation of RADA and detailed empirical results are publicly available for future intelligent tools that could automatically identify and migrate usage of deprecated RESTful API operations in client code.

\end{abstract}

\begin{IEEEkeywords}
API Deprecation, RESTful API, OpenAPI Specification, Web API, 
Evolution of Web APIs
\end{IEEEkeywords}

\section{Introduction}\label{sec.intro}
\input{introduction}

\section{Background}\label{sec.back}
\input{background}

\section{Empirical Study Setup }\label{sec.method}
\input{method}

\section{Results and Implications} \label{sec.results}
\input{result}

\section{Threats to Validity}\label{sec.threats}
\input{threats}

\section{Related Work}\label{sec.related}
\input{related}

\section{Conclusion and Future Work}\label{sec.con}
\input{conclusion}

\section*{Acknowledgement}\label{sec.ack}
\input{acknowledgement}



\balance
\bibliographystyle{IEEEtran}
\bibliography{ICSM20RestDep}
%

\end{document}

%% file: introduction.tex

Among all types of Application Programming Interfaces (APIs), web APIs play a crucial role as a pivotal interconnectivity mechanism to access software services over the Internet. Nowadays, web APIs are responsible for connecting software systems, data, and algorithms, enabling the creation of large software ecosystems by simplifying access to information and functionality~\cite{espinha2014web}. According to the world's largest web API directory ProgrammableWeb, the number of Web APIs has grown 1,000\% in the last decade from less than 2,000 listed in the directory in 2010 to over 23,000 in early 2020. 

Developers often create web APIs following the REST (REpresentational State Transfer) architectural style, which are referred to as RESTful APIs~\cite{fielding2000architectural}. Ideally, functions (HTTP requests) defined in RESTful APIs should not change, so that the API provider and consumers could evolve their service and software independently without affecting the other side. However, in reality, web APIs are often not static and even more change-prone than Java APIs adding new functionalities and fixing security bugs~\cite{li2013does}. Compared to non-web APIs, i.e., the APIs that do not need network connections, ever-evolving web APIs present unique challenges to stakeholders because it is often out of the control of the API consumers. For example, non-web API consumers can always use a local copy of the library (i.e. an older version). However, web API consumers have to modify the client application if the utilized service is no longer supported (e.g., services are shut down). 


Ideally, before being removed or modified, \textit{API elements} (e.g., class, method, and field in Java libraries), should be annotated as deprecated. Information such as replacement messages (e.g., using another API element) should also be provided to support developers to adapt the client application. Unfortunately, as revealed by prior studies~\cite{zhou2016api,nascimento2020javascript}, this deprecated-removed protocol is not always followed, introducing software maintenance issues for both API maintainers and consumers. Nevertheless, despite the increasing attention paid to the API deprecation in non-web APIs~\cite{zhou2016api,sawant2019react,brito2018use}, the deprecation practice in the web APIs or RESTful APIs has not yet been investigated. 


Our work aims to investigate how the deprecated-removed protocol is followed in RESTful APIs and characterize RESTful API deprecation practices. However, analyzing RESTful API deprecation is challenging due to three main reasons. First and foremost, unlike non-web APIs,  there is a lack of standard approaches for evolving RESTful APIs. Questions such as how to deprecate RESTful API and API elements and how to manage the RESTful API lifecycle remain unanswered. The Internet Engineering Task Force (IETF), a community in charge of the design of the internet, is still working on two new deprecation-related HTTP headers that would largely contribute to the standardization of web API deprecation~\cite{rfc,draftDeprecationHeader}. Secondly, limited resources are available for investigating the evolution of RESTful APIs, i.e., many APIs only provide the latest version of the API rather than the entire history. Last but not least, the documentation and deprecation-related information are often provided on APIs' official websites in custom-designed format. 

To achieve our goals, we conduct an empirical study on 2,224 versions of APIs from 1,368 RESTful APIs listed on APIs.guru. Our RESTful API set includes popular web APIs such as \textit{Gmail, Youtube, Amazon Elastic Compute Cloud (EC2), Instagram, Kubernetes, and Slack}. APIs.guru is the largest machine-readable profile directory for web APIs, and it is kept updated along with the evolution of web APIs~\cite{apisguru}. The collected API profiles from APIs.guru are in OpenAPI specification format, which is the standard language-agnostic interface to RESTful APIs~\cite{openapi}. OpenAPI specification allows both humans and computers to discover and understand the capabilities of the service(s) behind a RESTful API without access to its source code, documentation, or network traffic inspection. Furthermore, OpenAPI specifications resolve the challenge brought by customized API documentation and make large-scale exploration of RESTful APIs possible. 

We investigate the following research questions to guide our empirical study:
\begin{itemize}
    \item \rqone
    \item \rqtwo
    \item \rqthree
    \item \rqfour
\end{itemize}


To answer the above research questions, we first propose a new framework called RADA, which stands for RESTful API deprecation analyzer. RADA can identify deprecated RESTful \textit{API elements} (e.g., URI’s path, method, request parameters, responses arguments) in a given OpenAPI specification, and then summarize the deprecated API operations and their associated deprecation-related descriptions. We apply RADA on 2,224 collected OpenAPI specifications and perform our empirical study. Our empirical study finds that: i) out of 251 RESTful API versions that have introduced breaking changes upon the previous version, 87.3\% of them do not deprecate any operations in the previous version; ii) 38\% of the studied deprecation-related APIs contains more than 50\% operations impacted by deprecated API elements and 65\% of the impacted operations have deprecated request parameters; iii) 45\% of the studied deprecation-related APIs provide replacement messages for all the impacted API operations;  and iv) only 3 out of 219 studied deprecation-related RESTful APIs adopt a  proactive method for deprecation-related communication, i.e., using special HTTP headers and error code to inform developers when deprecated API operations are called in client code. 

To summarize, we make the following main contributions in this paper:
\begin{itemize}
    \item We design a framework called RADA that can automatically identify and analyze operations affected by deprecated API elements for RESTful APIs given OpenAPI specifications.
    \item We propose a heuristic-based approach for characterizing detailed deprecation information from an OpenAPI specification. 
    \item We conduct the first empirical study on the deprecation of the RESTful APIs stored in APIs.guru. The number of web APIs included in this work is the largest among existing empirical studies on web API evolution. We also provide important implications for RESTful API stakeholders to better facilitate ever-evolving web APIs.
    \item We provide a replication package\footnote{https://github.com/jerinyasmin/deprecation} with our preprocessed data and implementation of RADA in Python as a means to enable more in-depth studies on this topic.
 
\end{itemize}


%% file: background.tex
\subsection{RESTful API and RESTful API Deprecation}
REST is the most popular architectural design for building web services on top of HTTP~\cite{fielding2000architectural}. To access a resource (service) provided by a RESTful API, API consumers need to send an HTTP request, which is represented using a Uniform Resource Identifier (URI), to the server at API provider side and process responses returned by the server. Consider, for example, a service provided by GitHub v3 API to retrieve the list of repositories created by one GitHub user that is sorted by the creation time. This service can be accessed using the following URI:
\vspace{-0.16cm}
\begin{center}
\noindent{\footnotesize{
\textcolor{red}{GET} \textcolor{blue}{https://api.github.com/}\textcolor{orange}{users/$\langle$username$\rangle$/repos}\\ ?sort$=$created\&direcation$=$desc
}}
\end{center}
\vspace{-0.2cm}

\noindent The above URI is one specific service under an API operation named ``GET /users/$\langle$username$\rangle$/repos'' with its query parameter ``sort'' set to ``created'' and parameter ``direction'' set to ``desc''. http://api.github.com/ is the domain address. An \textit{API operation} in this work is referred to as a unique combination of HTTP method (e.g., GET) and a URI's path (e.g., /users/$\langle$username$\rangle$/repos). A request can be sent to the server with a set of request parameters associated with the API operation (e.g., query parameters and path parameters like 
username in the example), and a set of response arguments (e.g., error code) that defines the format of the response is returned by API provider's server.

Similar to non-web APIs, RESTful APIs follow the general lifecycle of software products, i.e., from creation and inception all the way to finally reaching their end-of-life. As mentioned in Section~\ref{sec.intro}, there is neither standard practice nor sophisticated support from HTTP for RESTful API deprecation. Some APIs such as \textit{Twitter Ads}, choose to deprecate the entire old version when breaking changes are bundled into a new version.\footnote{\url{https://developer.twitter.com/en/docs/ads/general/overview/versions}} Other APIs such as \textit{Kubernetes} choose to deprecate part of the services in the older version when a newer version introduces breaking changes.\footnote{\url{https://kubernetes.io/docs/reference/generated/kubernetes-api/v1.18}}

\subsection{OpenAPI Specification}
Several approaches have been proposed in industry towards an efficient and shareable formalism to describe RESTful APIs, e.g., WADL, RAML, API Blueprint, and OpenAPI Specification, formerly called Swagger~\cite{openapi-history}. In this work, we focus on OpenAPI specification, which is the format of API profiles on APIs.guru. OpenAPI specification is widely adopted in industry, and it is now maintained by OpenAPI Initiative (OAI) under the Linux Foundation. OpenAPI specification has two major versions, i.e., OpenAPI 3.0 and OpenAPI 2.0 (i.e., Swagger 2.0)~\cite{openapi}.

\begin{figure}[h]
\centering
\includegraphics[width=3.2in, keepaspectratio]{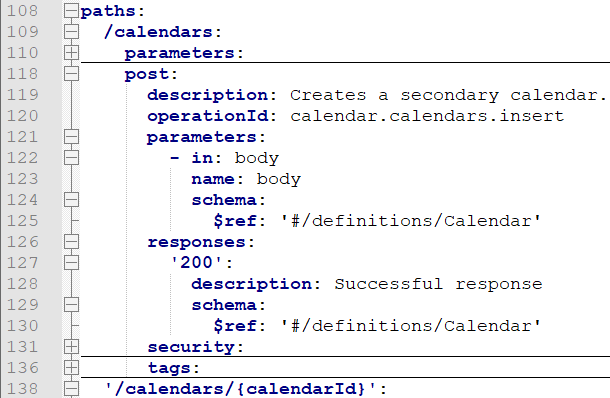}
\caption{OpenAPI 2.0 specification of a sample API. Same id represents the same API operation.}
\squeezeup
\label{fig:sample_spec}
\end{figure}

Figure~\ref{fig:sample_spec} shows part of the documentation of a sample API in our dataset in OpenAPI 2.0 specification format. In an OpenAPI specification, API elements are defined hierarchically in a JSON object. API operations are organized under an HTTP method object within a URI's path object. In Figure~\ref{fig:sample_spec}, the API operation “POST /calendars” is defined in the method object “post” (line 118-136) that is nested inside the “/calendars” path object (line 109). The shared parameters in each API operation under the same path are defined in a field named parameters (line 110-117). Each operation object contains the descriptions of all response arguments (line 126-130) and request parameters (line 121-125) that are associated with the operation.  A more detailed schema description can be found at OpenAPI’s homepage~\cite{openapi}. 
\vspace{-0.2cm}
\vspace{0.05cm}

%% file: method.tex

\subsection{Data Collection}
\label{sec:dataCollection}
\begin{figure}
    \centering
    \includegraphics[width=3.4in, keepaspectratio]{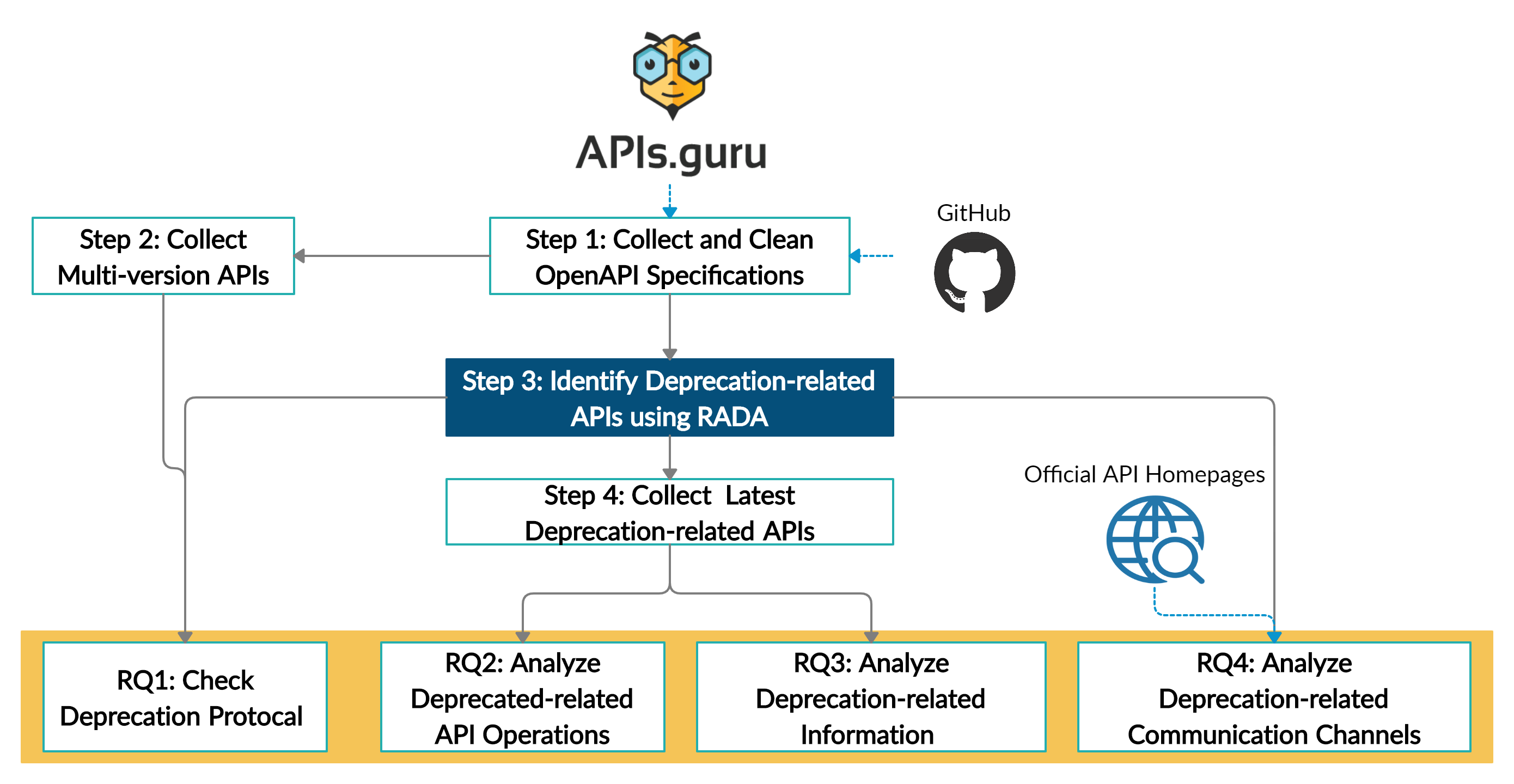}
    \caption{An overview of the data collection process. Dashed lines indicate a connection to a data resource. Process box in blue represents steps requiring RADA, our proposed RESTful API deprecation analyzer.}
    \squeezeup
    \label{fig:dataCollection}
\end{figure}

We collected target RESTful APIs for our empirical study from APIs.guru, the “Wikipedia” of web APIs. APIs.guru was launched in 2016 to host machine-readable documents (i.e., OpenAPI 2.0 and 3.0 specification) for publicly available RESTful APIs. We selected APIs.guru mainly due to three reasons: i) it has the most comprehensive and up-to-date list of OpenAPI specifications provided either directly by public RESTful APIs or transferred from official API documentation by domain experts; ii) it stores historical OpenAPI specifications of their collected RESTful APIs on GitHub, and thus makes the analysis of RESTful API evolution possible; iii) it has been considered as the proxy of RESTful API documentation in prior studies on web APIs~\cite{d2spec,wittern2017opportunities}. In the following, we describe how we collect specific pieces of data on which we relied to answer the four RQs raised in Section 1. Figure~\ref{fig:dataCollection} presents an overview of the data collection process. 

\noindent\textbf{Step 1. Collect and clean OpenAPI specifications.} We downloaded the OpenAPI-directory repository created by APIs.guru from GitHub in October 2019. We then extracted all OpenAPI specifications by searching files named “swagger.yaml” (following OpenAPI 2.0 specification) or “opeapi.yaml" (following OpenAPI 3.0 specification). In total, we collected 3,536 specifications from 1,595 unique RESTful APIs. The number of specifications is larger than the number of unique APIs because some APIs have multiple specifications corresponding to multiple release versions. Next, we applied the following two filters to clean the collected OpenAPI specifications.
\begin{itemize}
    \item Filter 1: Removing redundant and erroneous OpenAPI specifications. Four APIs were removed as they are duplicate with existing OpenAPI specifications. Another four OpenAPI specifications have YAML errors inside the file. Thus they were excluded from our dataset as well. 
    \item Filter 2: Removing specifications from non-stable released versions. We only considered the OpenAPI specifications from stable release versions as unstable API versions are usually maintained separately from stable versions and contain more experimental changes. Specifically, we excluded the versions that have the following keywords (case-insensitive) in their version name: ``alpha'', ``beta'', ``preview'' (also ``-pre''), and ``sandbox''. 
\end{itemize}


\noindent\textbf{Step 2. Collect multi-version APIs.} To answer RQ1, we extracted breaking changes introduced in each API version by comparing its specification to the one from its previous version. Thus we identified APIs with multiple specifications in our dataset and created a multi-version API corpus. This corpus became the first information source for answering RQ1. 

\noindent\textbf{Step 3. Identify deprecation-related API versions using RADA.} All RQs rely on an approach that can identify and summarize all API operations that are impacted by deprecated API elements in a given OpenAPI specification. This approach is generalized as a new framework, i.e., RADA, which is presented in Section 3.2. We applied RADA on all the 2,224 collected OpenAPI specifications and found 257 (11.55\%) \textit{deprecation-related API versions}, i.e., the specifications contain at least one deprecated API element. These 257 versions are from 219 unique APIs. The results of RADA for each API specification became the second information source for answering RQ1. In addition, the 219 deprecation-related APIs became the target APIs for RQ4 where we manually investigate communication channels for deprecation notification. 

\noindent\textbf{Step 4. Collect the latest deprecation-related APIs.} In RQ2-4, we focus on the latest deprecation-related API version of each target API that has at least one deprecation-related version. This means that if a target RESTful API has more than one deprecation-related version, we only kept the latest version. We designed this mechanism to capture the latest deprecation practices in RESTful APIs. In the end, we collected 219 deprecation-related API specifications, each representing one unique RESTful API. These API specifications were the source of information for addressing RQ2 and RQ3. 
We summarize the basic statistics of our target data set for answering four RQs in Table~\ref{tab:statistics}. 

\vspace{-0.2cm}
\begin{table}[h]
    \centering
    \caption{The statistics of the collected data set.\label{tab:statistics}}
    \begin{tabular}{p{0.35\textwidth}p{0.08\textwidth}}
    \toprule
    {\em Step 1. Collect and clean OpenAPI specifications} & \\
        Number of OpenAPI Specifications &  \multicolumn{1}{r}{2,224}\\
         Number of Unique RESTful APIs & \multicolumn{1}{r}{1,368} \\
         \quad OpenAPI Version 2.0 & \multicolumn{1}{r}{1,290}\\
         \quad OpenAPI Version 3.0 & \multicolumn{1}{r}{78}\\
         \hline
    {\em Step 2-4. Multi-version and deprecation-related APIs (RQ1-4). } & \\
         Number of multi-version APIs & \multicolumn{1}{r}{212} \\
         Number of specification  of the multiple-version APIs & \multicolumn{1}{r}{1,068}\\
         Number of deprecation-related APIs & \multicolumn{1}{r}{219}\\
         Number of deprecation-related API versions & \multicolumn{1}{r}{257}\\
         \hline
         {\em Examples of Notable web APIs} &\\
         \multicolumn{2}{p{0.4\textwidth}}{Gmail, Youtube, Amazon Elastic Compute Cloud (EC2), Instagram, Github, Jira, Kubernetes, Slack}\\
                  \bottomrule

    \end{tabular}
\vspace{0.1cm} 
 \squeezeup 
\end{table}
\subsection{RADA: RESTful API Deprecation Analyzer}
\label{sec:rada}
RADA takes as input an OpenAPI specification and then outputs a set of API operations with their associated deprecation-related information. If any API element inside an API operation gets deprecated, the API operation is labeled as \textit{deprecation-related API operation}. RADA is designed based on our observation that RESTful API providers often use the optional “deprecated” field or deprecation keywords, i.e., any words containing the prefix “deprecat” (e.g., deprecated, deprecation, and deprecating), to specify deprecated API elements. RADA identifies deprecated API elements and versions using the following three deprecation criteria:
\begin{enumerate}
    \item An API element is deprecated if the optional boolean type field ``deprecated” is set to ``true”.
    \item An API element is deprecated if the description (e.g., in the ``summary" or ``description" field) of the element contains at least one deprecation keyword.
    \item An API version is fully deprecated if the value of the optional string type field ``description” inside the InfoObject contains at least one deprecation keyword.
\end{enumerate}

Note that RADA lowercases all text before checking deprecation keywords. It also excludes the following cases for criterion 2-3:
\begin{itemize}
    \item The name of the target request parameter or response argument contains deprecation keywords. For instance, \\
\textit{\textbf{deprecated}ApiVersion: `Earliest RE API version supported, including \textbf{deprecated} versions.'}
    contains the keyword ``deprecated'' in the description, but the keyword is describing the function of the field.
\item The deprecation keyword is inside single/double quotation marks.
For instance,
\textit{
   `Tags for this dimension.
   Examples: "default", "preview", \textbf{``deprecated"}.}
contains the keyword ``deprecated”, but it represents a sample value for the field.
\end{itemize}
\vspace{-0.2cm}
\begin{algorithm}
\caption{Identifying and analyzing deprecations in \\OpenAPI specification.}\label{alg:rada}
\scriptsize
\begin{algorithmic}[1]
\Procedure{RADA}{$s$}
\State $results\gets \{\}$
\State $isDpr \gets checkDeprecation(s.infoObj)$
\State $globalDef\gets s.globalDefinitions$
\ForAll{$path \in s.paths$}
    
    \ForAll{$method \in path$}
        \State $opNm\gets method.name + path.name$
        \State$hasOp,txtOp\gets checkDeprecation(method)$
        \State$respObjs,reqObjs\gets getObjs(method,globalDef)$
        \ForAll{$respObj \in respObjs$}
            \State$hasResp, respTxt \gets checkDeprecation(respObj)$
        \EndFor
        \ForAll{$reqObj \in reqObjs$}
            \State$hasReq, reqTxt \gets checkDeprecation(reqObj)$
        \EndFor
        \If{$hasOp$ or $hasResp$ or $hasReq$}
        \State$isDpr \gets true$
        \EndIf
        \State{$results.put(opNm,\{isDpr,hasOp,hasResp, hasReq, opTxt,$} 
        \State{$ respTxt ,reqTxt \})$}
    \EndFor

\EndFor
\State \textbf{return} $results$
\EndProcedure

\end{algorithmic}
\end{algorithm}
\vspace{-0.3cm}
Algorithm~\ref{alg:rada} shows the main process of RADA. It reads an API specification  {\em s} as input, and outputs a map results containing each API operation  {\em opNm} and its associated seven pieces of deprecation-related information, i.e.,  {\em isDpr},  {\em hasOp},  {\em hasResp},  {\em asReq},  {\em opText},  {\em respText}, and  {\em reqText}. The first four are boolean variables representing if {\em opNm} is deprecation-related, having deprecated operation(s), having deprecated response argument(s), and having deprecated request parameter(s), respectively. The last three string variables refer to the text containing deprecation keywords in the description/summary of deprecated operation, response arguments, and request parameters.

RADA parses the input specification  {\em s} and determines if the entire API version is deprecated following deprecation criterion 3 (line 3). It then retrieves the definitions of common data structures {\em globalDef} from the global section in {\em s} (line 4). This variable is required to generate the complete request objects and response objects within each API operation (line 9). Next, RADA processes each API operation by visiting each method object inside each path object (line 5-14). {\em opNm }is determined by each visited method and its outer path object (line 7). For each operation, RADA determines if the operation itself is deprecated (line 5-8) and extracts deprecation-related text (line 8) following deprecation criterion 1 and 2. This process is then repeatedly applied to each request and response object (line 10-11, line 13-14). After collecting all deprecation-related information from operation, {\em reqObj}, and {\em respObj}, RADA will set the deprecation-related variable isDpr of {\em opNm }as true if opNm contains at least one deprecated element (line 16-17). For non deprecation-related API operations, {\em isDpr}, {\em hasOp}, {\em hasResp}, {\em hasReq} are set to false, while {\em txtOp}, {\em txtResp}, and {\em txtReq} are set to null. 

\noindent\underline{\em Evaluating RADA.} 
RADA is built upon a set of heuristic keyword-based rules. Thus it might suffer from some false positive and false negative cases. False positive cases represent non-deprecated API elements that are identified as deprecated because it contains at least one deprecation keyword. False negative cases represent deprecated API elements that RADA missed as it does not mention any deprecation keyword. To further migrate this threat to validity, we randomly sampled 50 API versions that contain 3,444 API operations. We carefully read the definitions associated with each sampled API operations and manually identified if the operation is deprecated or not. We then compared the labels returned by RADA with our manually created labels. Our evaluation results show that at API version level, RADA achieves an accuracy of 100\%, i.e., all API versions that contain at least one deprecated element is captured, and all captured API versions indeed contain deprecated elements. At API operation level, RADA achieves a precision of 94\% and a recall of 100\%. 


\vspace{0.1cm}

%% file: result.tex
\subsection{\rqone}
\input{rq1}

\subsection{\rqtwo}
\input{rq2}

\subsection{\rqthree}
\input{rq3}

\subsection{\rqfour}
\input{rq4}

\subsection{Implications}
Our empirical study results reveal critical issues in RESTful API deprecation. These issues provide opportunities for future research to better support the web API ecosystem. 

\noindent\textbf{Guidelines for a clear and consistent RESTful API deprecation policy.} Our study (RQ1) shows that almost half of the studied RESTful APIs do not follow the deprecated-removed protocol, i.e., breaking changes are introduced without alarming developers. Also we observe that even within one API and organization, there exist inconsistent deprecation behaviors, e.g., some Azure APIs follow the deprecated-removed policy and some do not. 
Given the increasing need to evolve RESTful APIs, the lack of a clear and consistent deprecation model causes maintenance challenges for API consumers. it is necessary for API maintenance teams and organizations to set clear RESTful API deprecation policy for their APIs and follow the agreed policy persistently. In the deprecation policy, API providers should provide information such as how the organization will deprecate API elements or versions, when will deprecated APIs be removed and turned into sunset mode, what deprecation information to provide, and how developers will be informed and communicated in terms of deprecation-related issues, etc.

\noindent\textbf{Tools for automatic detection of deprecated API use in client applications.} As shown in RQ4, most of the deprecation information is communicated in a reactive way, e.g., in official websites. Proactive approaches can better keep API consumers informed of the deprecated APIs in the client code given the frequent and non-trivial deprecation as our RQ2 reveals. Automated tools are needed to automatically analyze API documentation and keep track of deprecation impacted operations. Yang et al. ~\cite{d2spec} took a first attempt towards generating OpenAPI specifications from document. The up-to-date specifications can then be leveraged by automatic tools in detecting usage of deprecated APIs in client code. Our proposed RADA can contribute to such effort by identifying deprecation-related operations from OpenAPI specifications. 

\noindent\textbf{Tools for migration support on deprecated web APIs.}
Our study shows that only 45\% of studied RESTful APIs provided replacement messages for deprecated API elements (RQ3). Without such recommendations, API consumers may face challenges when migrating the deprecated APIs.
There are possible reasons behind the lack of replacement messages. First, there might be no replacement plan for the deprecated API elements. In such cases, we suggest the API providers follow the standard deprecated-removed model commonly adopted in the ecosystems such as Android~\cite{zhou2016api} so that developers can have sufficient time to prepare for final removal of the APIs. The second reason is that API providers may forget to mention replacement in the documentation. In such cases, in addition to suggest the API providers to keep their document up-to-date with deprecation information, we believe automated tools should be proposed to automatically mine alternative API elements for deprecated APIs and recommend them to API consumers, similar to prior studies in Android and Java~\cite{A3, 10.1145/2000799.2000805}. Furthermore, such recommendation tools can be combined with tools that detect deprecated API use in client code to automatically update client code.


\noindent\textbf{Well-maintained repositories of web API artifacts.} Although our study includes the largest collection of web APIs in literature, the scale is relatively smaller compared to existing studies on other non-web APIs (e.g., Android~\cite{ li2018characterising}, Java~\cite{xavier2017historical}
), which have well-maintained large-scale repositories. 
This also brings challenges for API consumers who may need to figure out how to migrate to the latest version and would need to compare the artifacts (e.g., document, or OpenAPI specifications) of different versions. Thus we encourage more effort to adopt and maintain OpenAPI specifications from API providers, and more community effort (e.g., APIs.guru) to curate and maintain repositories to store web API artifacts. 

%% file: rq1.tex
\noindent{\bf Motivation.} As revealed by existing studies on the deprecation of non-web APIs~\cite{zhou2016api}, API providers may not follow the deprecated-removed protocol, i.e., they directly introduce breaking changes without leaving enough transition time for API consumers. In this paper, \textit{breaking changes} refer to updating or removing API elements, including operations, request parameters, and response arguments. In RESTful APIs, where the web API consumers have no control over the API and the service behind the API, the impact of not following the deprecation protocol can be even more severe than in non-web APIs. Hence, in the first research question, we investigate how the deprecated-removed protocol is followed in practice.

\noindent{\bf Approach.} To answer RQ1, we start with multi-version web APIs that have multiple versions stored in APIs-guru. The prevalence of multiple versions of the same APIs allows us to track the evolution of the APIs and understand what breaking changes have been introduced during evolution. More specifically, in Section~\ref{sec:dataCollection}, we observed that only 16\% of the 1,368 target RESTful APIs contain deprecation-related versions. But it does not mean that the remaining 1,149 APIs do not follow the deprecated-removed protocol. Because many of the target APIs may have only one released version or do not need to deprecate any API elements as they always guarantee backward compatibility across versions. Therefore, RQ1 targets the multi-version APIs that have introduced breaking changes as compared to previous releases, i.e., they are the ones needed to follow the deprecated-removed protocol. 

For a multi-version API, we first sort the specifications of the multiple versions by release date or version id in ascending order, e.g., \{{\footnotesize\ttfamily V1, V2, V3}\}. Then starting from the second collected version of the API (i.e., {\footnotesize\ttfamily V2}), we extract the breaking changes introduced by comparing its specification with that of the previous version, i.e., ({\footnotesize\ttfamily V2, V1}), and ({\footnotesize\ttfamily V3, V2}). We adapt a popular OpenAPI specification diff tool, named {\em SwaggerDiff}~\cite{swaggerdiff}, to perform the comparison between two consecutive specifications. As {\em SwaggerDiff} only supports OpenAPI 2.0, we convert the 78 OpenAPI 3.0 specifications to 2.0 using a tool named api-spec-converter~\cite{lucybot}, to perform the comparison between two consecutive specifications and identify breaking changes.

Next, we apply RADA on the relevant API specifications to identify whether they are deprecation-related (i.e., mention deprecation-related information) before introducing breaking changes. For instance, if we identify that {\em V2} introduces breaking changes over the previous version {\em V1}, we apply RADA on {\em V1} to decide whether {\em V1} has any deprecated API element. Note that we did not check if the deprecated API elements match with the breaking changes, as we focus on understanding whether one API version adopts deprecation practice or not, rather than finding the individual API elements that API providers fail to mark as deprecated. Hence results from our study will be an upper-bound for the number of APIs following the deprecated-removed protocol. In practice, the number of APIs fully following the protocol, i.e., annotating each deprecated API element, would be even less than the one we reported.

\begin{figure}[t]
\captionsetup{skip=0pt}
\centering 
\includegraphics[width=3.3in, height=1.6in,keepaspectratio]{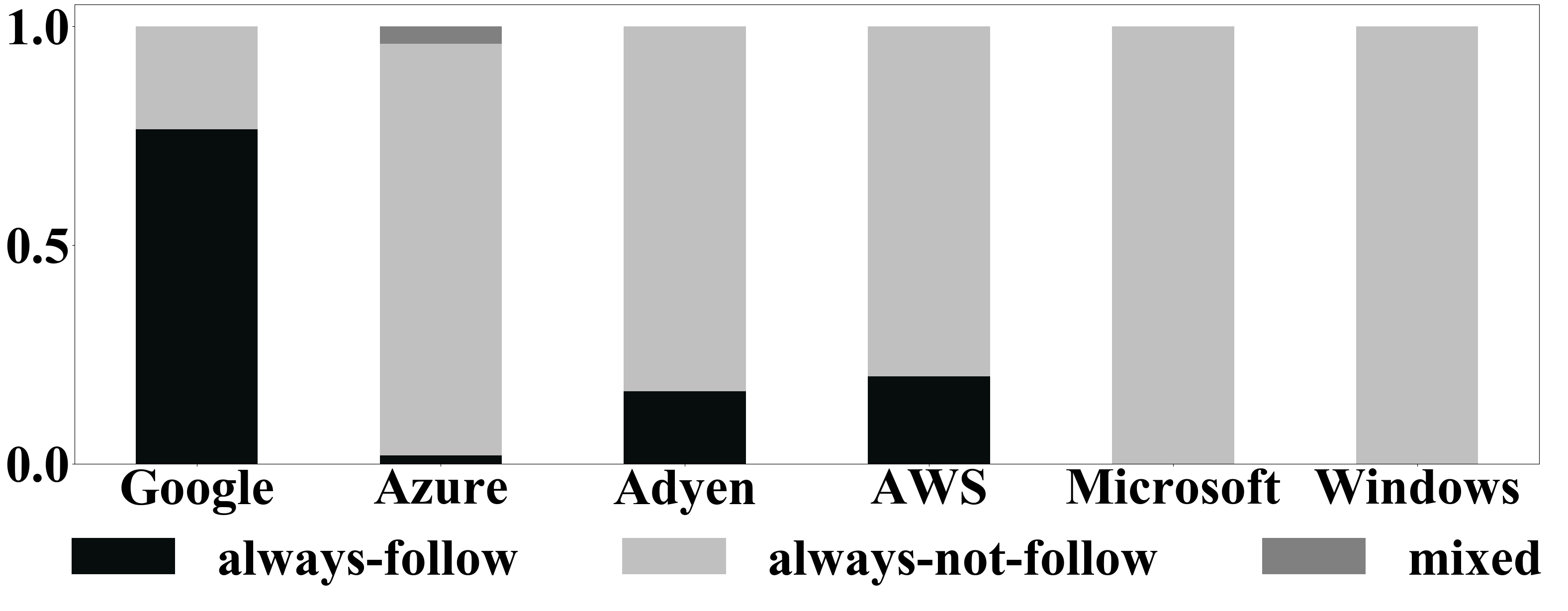}
\caption{Deprecation behavior of APIs in each organization}
\squeezeup
\label{organization_behavior}
\end{figure}

\noindent\textbf{Results.} We applied the adapted SwaggerDiff on 1,068 specifications from 212 multi-version APIs. Two specifications failed to be parsed. In the end, we identified 251 versions that have introduced breaking changes upon the previous version. Among them, for only 32 versions (12.7\%), their previous API versions are deprecation-related, while the remaining 219 (87.3\%) API versions did not specify any information related to deprecation before making changes. 
We then went further to investigate how each API behaves regarding the deprecated-removed protocol. We defined three types of behaviors at API level: i) always-follow, for all versions that introduced breaking changes, its previous version is deprecation-related; ii) always-not-follow, for all breaking change introducing versions, the previous version does not provide any deprecation information; iii) mixed, some versions follow and some do not. We found that out of 133 considered APIs, only 16 of them always follow the protocol, four have mixed behavior, and the remaining 113 (84.3\%) always do not follow. 
We performed a similar study at organization level. The 133 APIs come from six organizations, i.e, Azure (102), Google (17), Adyen (6), AWS (5), Microsoft (2), and Windows (1). For each organization, we calculate the number of APIs under the organization for each of the three deprecation behavior types and plot the results in Figure~\ref{organization_behavior}. We can observe that even within the same organization, deprecation behaviors can vary cross APIs. Google did the best among all considered organizations with 69\% (9/13) of their APIs always follow the deprecated-removed protocol. The four mixed behavior APIs are all from Azure. 
As the above results reveal the lack of standard policy on following the deprecated-removed protocol in RESTful APIs, we also performed an initial investigation on potential reasons for not deprecating breaking changes. We hypothesized that if there are only few operations impacted by breaking changes in one version, developers might choose to ignore the deprecated-removed protocol as they may assume few changes will not affect many API consumers. To test this hypothesis, we compared the number of API operations impacted by removed/modified API elements between API versions following the deprecation protocol and the ones that do not follow. 
We observed that indeed API versions do not follow the deprecation protocol have much less impacted operations by breaking changes than the ones following the protocol. In fact, 73.5\% (161/219) of the API versions that do not follow deprecation have less than 10 operations impacted by breaking changes. The same ratio is 47\% (15/32) for API versions following deprecated-remove protocol. The median number of impacted operations by breaking changes in APIs follow and not follow the deprecation protocol is 10 and 5 respectively.
\vspace{0.05cm}
\begin{tcolorbox}
RQ1-Findings: 87.3\% of RESTful API versions that introduced breaking changes do not provide any deprecation information in the previous version. Within each API and organization, various deprecation behaviors are observed, i.e., some versions follow the deprecated-removed protocol while others do not. Versions with more impacted operations by breaking changes are more likely to mention deprecation-related information in the previous version.
\end{tcolorbox}
\vspace{-0.4cm}

%% file: rq2.tex
\noindent\textbf{Motivation.} There exists no standard way to deprecate RESTful APIs, which means that API providers will adopt different deprecation strategies. In the second research question, we categorize patterns in the deprecation of RESTful APIs. Such knowledge would help API providers understand the practice and help API consumers evaluate the potential risk of adopting a particular deprecation-related version. 

\noindent\textbf{Approach.} We characterize deprecation patterns in RESTful APIs from three aspects, i.e., the prevalence, method types, and the deprecated source of impacted API operations. RQ2 targets on the lastest 219 deprecation-related versions (ref. Section~\ref{sec.method}). We describe the detailed methodology for each considered aspect as follows.

\begin{itemize}
    \item \textbf{Prevalence:} Estimating the prevalence of deprecated API operations is straightforward with the output returned by RADA. We calculate the number of total API operations provided in each target API, and the number of API operations that are identified as “deprecation-related” by RADA. The deprecation-related operation ratio (\#deprecation-related operations/\#total operations) is then calculated across the target set. 
    \item \textbf{Method Type:} RESTful API operations can be categorized based on their HTTP method type. Based on the information returned by RADA for each target API, we first calculate the total number of API operations for each type of methods, and then calculate the number and ratio of operations in each method type that are deprecated. 
    \item \textbf{Deprecated Source:} A deprecation-related API operation might be impacted by deprecated operation, request parameters, or response arguments. Based on the results returned by RADA, we calculate the ratio of deprecation-related API operations that have deprecated operation, request parameter(s), and response argument(s), respectively. Note that one deprecation-related operation could have multiple deprecated sources. 
\end{itemize}

\noindent\textbf{Results.} We find that, on average, 46\% of the operations in each studied API are deprecation-related. We then create five groups of APIs based on their ratio of deprecation-related operations: (0, 25\%), [25\%, 50\%), [50\%, 75\%), [75\%. 100\%), and 100\%. The most popular group is (0, 25\%), which means most of the studied APIs have less than 25\% total operations affected by deprecated API elements. Few APIs have more than 25\% and less than 100\% deprecation-related operations: group [25\%, 50\%), [50\%, 75\%), [75\%. 100\%) contains 25, 18, and 3 APIs, respectively. Surprisingly, we also find that 71 out of the 219 studied APIs have all operations affected by deprecated API elements. We then manually investigate the APIs in this group and find that: i) four APIs specified in the info section that the entire API version is deprecated and API consumers are suggested to move to the latest version of the API; ii) many APIs in this group only have few operations, thus it is much easier for them to be impacted at the same time; iii) some shared (mostly query) parameters within/cross path object get deprecated and they affect all operations that can be called together with the parameters.



\begin{figure}[t]
\centering
\captionsetup{skip=0pt}
\includegraphics[width=3.4in,
  height=1.6in,
  keepaspectratio,]{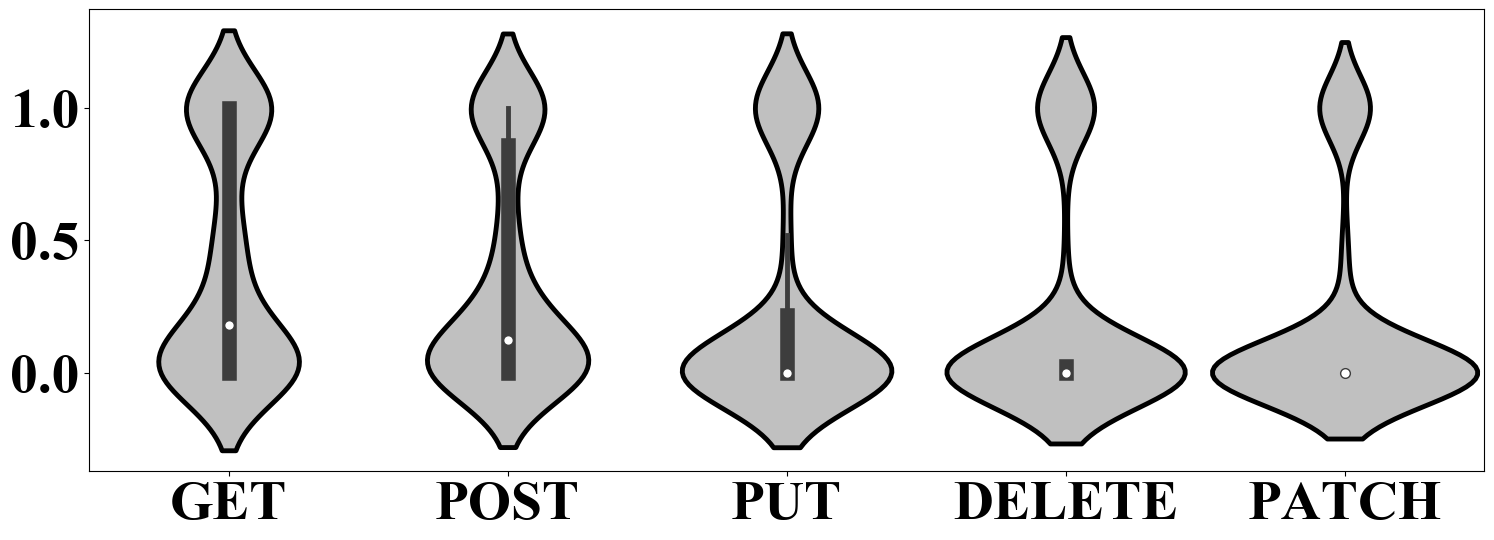}
\caption{Ratio of API operations impacted in five types of methods}\label{fig:rq2.2}
\label{impacted_methods}
\squeezeup
\vspace{-0.05cm}
\end{figure}
\begin{figure}[h]
\captionsetup{skip=0pt}
\centering
\vspace{-0.1cm}
\includegraphics[width=3.4in,
  height=1.6in,
  keepaspectratio]{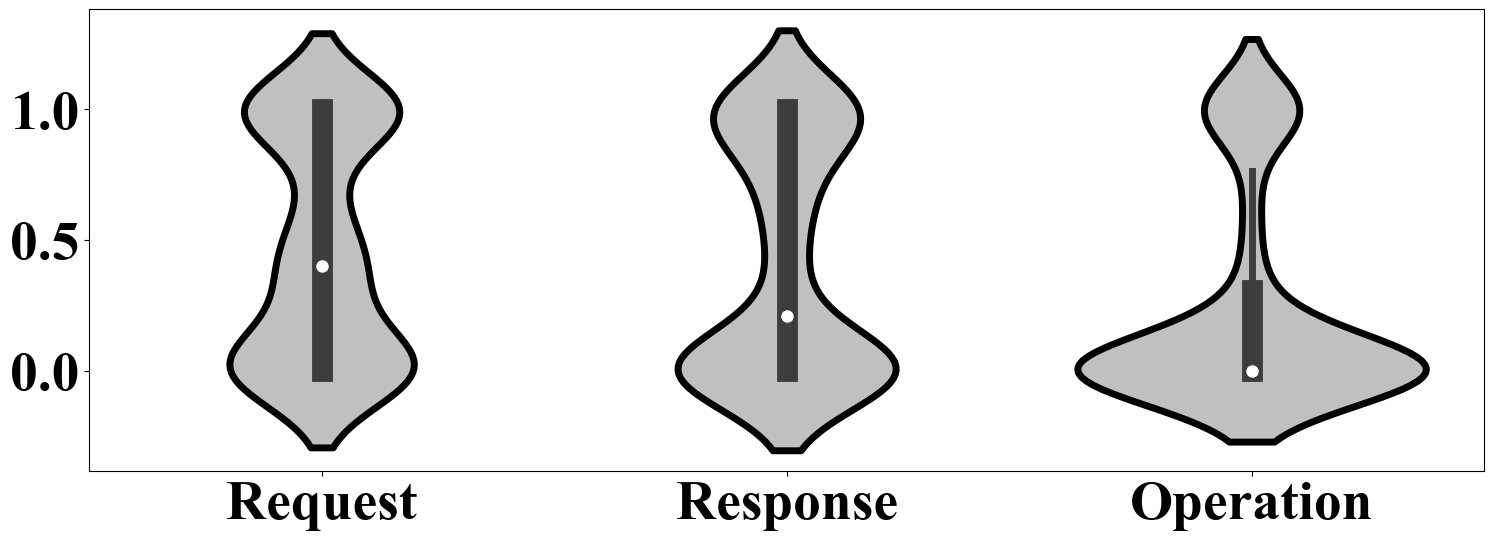}
\caption{Ratio of API operations impacted by three sources}\label{fig:rq2.3}
\label{impacted_operations_types}
\end{figure}

Figure~\ref{fig:rq2.2} presents the violin plots for different HTTP methods based on their deprecation-related operation ratios in the studied 219 APIs. We find that GET operations are more likely to be deprecation-related, followed by the POST operations. Specifically, 71 out of 219 APIs contain deprecation-related GET operation(s), and in 65 APIs, all of their GET operations are affected by deprecated API elements. On average, 41\%, 35\%, 25\%, 20\%, and 17\% of the GET, POST, PUT, DELETE, and PATCH operations are deprecation-related in each target API. The median of impacted API operation ratio for GET and POST operations is 18\% and 12\% respectively, while the median for the other three types of operations is 0\%.
Figure~\ref{fig:rq2.3} presents the violin plots for three deprecation sources based on their deprecation-related operation ratios in each target API. We can observe that most deprecation-related operations are impacted by deprecated request parameters, followed by deprecated response arguments. A fewer number of operations have directly deprecated operations. On average, 46\%, 42\%, and 24\% of deprecation-related API operations contain deprecated request parameters, response arguments, and operation, respectively with a median of 40\%, 21\% and 0\%.
\vspace{-0.2cm}

\begin{tcolorbox}
RQ2-Findings: On average, 46\% of the operations within a target RESTFul API are deprecation-related. The ratio of deprecation-related operations varies among APIs: all operations in 71 out of 219 target APIs are affected by deprecated API elements, while 102 APIs have less than 25\% API operations affected. GET operations are more likely to be deprecation-related. Most deprecation-related operations (46\%) have deprecated request parameters.
\end{tcolorbox}

%% file: rq3.tex
\noindent{\bf Motivation.} API providers should always provide clear deprecation-related information, including replacement messages (i.e., which operation could be used to implement the same function), when the operation will be removed, etc. Insufficient deprecation information leaves API consumers clueless on planning what they should do to preserve the existing functionalities when the APIs in use are deprecating. Lack of sufficient deprecation information is a known problem in local APIs, e.g., 21.4\% of the deprecated methods in the latest Android release (level 28) do not have replacement messages~\cite{li2018characterising}.
However, it is unknown to what extent deprecation information is provided for RESTful APIs. 
Thus in this RQ, we investigate the details of the deprecation information provided in OpenAPI specifications.

\noindent{\bf Approach.} We performed a semi-automatic analysis on the textual descriptions of relevant fields in OpenAPI specifications and characterized what types of deprecation information are provided. In particular, we analyzed the latest versions of the 219 deprecation-related web APIs, which are detected by RADA. In total, RADA reports 880 textual descriptions (i.e., from ``description'' or ``summary'' in OpenAPI specification) that contain deprecation-related keywords (Section~\ref{sec:rada}). The detected 797 textual descriptions are from a total of 203 web APIs. Note that some API elements are marked as deprecated (i.e., the deprecation field is set to be {\em true}) however do not contain any textual descriptions, thus we excluded such API elements in this RQ. 
We characterized and summarized the deprecation information for RESTful APIs following open card sorting  method~\cite{spencer2004card}. More specifically, we applied open card sorting on a randomly-selected set of 100 textual descriptions and summarized a total of five types of deprecation information. Then we designed a heuristic-based approach to automatically characterize the deprecation information in the remaining 697 (i.e., 797-100) textual descriptions. 

Below we describe the details of each type of deprecation information (namely replacement message, external reference, explanation, deprecation, and removal time) and how the heuristic-based approach works to automatically detect them. Specifically, we processed each textual description (i.e., removing extra spaces and converting to lowercase) and applied the following rules to decide whether one type of deprecation information is described in the textual description.


\noindent\textbullet{} {\em Replacement Message.} Alternatives are recommended to replace the deprecating API elements. An example of recommending an alternative parameter is ``{\em deprecated in favor of desired\_bundle\_size\_bytes}''. We concluded common keywords used to indicate replacement and leverage them to detect whether one textual description lists replacements, i.e., ``deprecated by", ``recommend'', ``use '', ``in favor of'', ``replaced by", ``instead", ``merged into'', and ``see and preferred”.

\noindent\textbullet{} {\em External Reference.} API providers may include detailed deprecation information in external documents, such as ``{\em this group version of deployment is deprecated by apps/v1/deployment. see the release notes for more information}''. We concluded the common types of external documents and use them in the detection of external references, i.e., ``release note'', ``changelog'', ``migration''.

\noindent\textbullet{} {\em Explanation.} API providers may explain the reasons behind deprecation, such as ``{\em  this  property  has  been  deprecated  due  to  privacy  changes}''. We search for keywords, i.e., ``due to”, to identify the reason for deprecation.

\noindent\textbullet{} {\em Deprecation Time.} The time/version when one API element starts to deprecate may be specified, such as ``{\em  this  field  has  been  deprecated  as  of  version  2.1}''.
We use whether or not containing the keyword “deprecated in” for detecting deprecation time. We exclude the cases having ``deprecated in favor” due to its frequent use to express replacement message.

\noindent\textbullet{} {\em Removal Time.} API providers may indicate the official removal time of one deprecating API element so the consumers can better plan for it, such as ``{\em deprecated and will be removed in gitlab 9.0.}''.
We use the following keywords
to detect the removal time/version of one API element, i.e., ``removed in” , ``next release”, ``next version”, and ``next api”.
\begin{table}[t]
    \centering
    \caption{The performance of the heuristic-based approach in characterizing types of deprecation information}
    \begin{tabular}{lrr }
    \hline
     \textbf{Type} & \hfil\textbf{Precision} & \hfil\textbf{Recall}  \\
    \hline
    Replacement Message  & 96\% & 100\%  \\  
        External Reference &   81\% & 100\% \\ 
    Explanation & 100\% & 75\% \\  
    Deprecation Time & 83\% & 46\% \\ 
    Removal Time &  100\% & 86\% \\ 
    \hline
    \end{tabular}
    \squeezeup

    \label{tab:detection_results}
\end{table}
\noindent\underline{\em Evaluating the detection of deprecation information.} 
We took a statistically significant (95 $\pm$ 5\%) sample of 248 textual descriptions for this evaluation. The population is 697 (i.e., 797-100) textual descriptions. We manually examined each of the textual descriptions and decided whether it contains one or more of the five types of deprecation information. The manually-obtained ground-truth is then compared with the results of the heuristic-based approach to evaluate the performance, i.e., measured by precision and recall (Table~\ref{tab:detection_results}). Precision shows how many of the detected instances are correct in percentage (i.e., consistent with the ground-truth) for each type. Recall shows the percentage of the ground-truth instances that can be detected correctly.
Overall the precision and recall for each are acceptable except for deprecation time, for which the recall is not high (46\%). We excluded deprecation time from further analysis.



\noindent{\bf Results.} We summarized that per studied API, how many of its deprecation-related operations provide any of the five types of deprecation information. Similar to the previous RQs, we concluded at the operation level, i.e., if any of the API element of one operation contains one type of deprecation information, this deprecation-related operation provides this type of deprecation information. For each of the four types, we calculated a ratio as the percentage of the deprecation-related operations with this type of deprecation information out of all the deprecation-related operations in one API.

For each type of deprecation information, e.g., replacement message, we obtain a set of ratio values (one set per API). By analyzing the set of ratio values for replacement message, we find that 
only 45.20\% (99/219) of the APIs provide replacement suggestions for all their deprecation-related operations. On the contrary, we find that 33.3\% (73/219) of the studied web APIs do not suggest replacements for any of the deprecation-related operations. The remaining APIs (21.6\%) suggest alternatives {\em occasionally} for some of their deprecation-related operations. 

Interestingly, our study shows that the replacement message is the most frequently-specified deprecation information across the studied web APIs. For the other three types of deprecation information, the majority of the APIs do not provide them at all, i.e., 214 APIs never provide removal time, 208 for external references, and 215 for an explanation. The detailed distributions of the three types are omitted due to space constraints.

\begin{tcolorbox}
RQ3-Findings: 73 of the 219 (33.3\%) studied RESTful APIs do not provide any replacement messages. 
Only 45\% of the studied APIs provide replacement messages for all the deprecation-related operations, which is much lower compared to Java and C\# APIs (66.7\% and 77.8\% respectively in the literature~\cite{brito2018use}). The other types of deprecation information, such as removal time, are rarely provided, e.g., only 5\% of deprecation-related APIs mention removal time. 
\end{tcolorbox}
\vspace{-0.3cm}

%% file: rq4.tex
\noindent{\bf Motivation.} Communicating deprecated API elements is an important activity to guarantee that API consumers are aware of the deprecated elements which might affect their code. In non-web APIs, communication can be easily done via support from programming languages. For instance, since J2SE 5.0, Java provides a mechanism to deprecate API elements including types, methods, and fields, using the ``@Deprecated'' annotation. This annotation causes the compiler to raise a warning message when it finds that the deprecated API elements are in use. Java also provides a ``@deprecated'' Javadoc tag that can make Javadoc show a program element as deprecated. However, web APIs including RESTful APIs do not have similar support from HTTP yet~\cite{draftDeprecationHeader}. Thus in the final research question, we investigate what are the additional communication channels leveraged by API providers to communicate with API consumers. 

\noindent{\bf Approach.} We performed a manual inspection on the official websites of the 219 deprecation-related APIs identified by RADA (Section~\ref{sec.method}) and find two types of communication channels. The first type is called \textit{technical communication}. A sample technical communication channel is the API response. For instance, some web APIs can use custom HTTP headers (e.g., ``X-API-Warn'' in  \textit{Clearbit} API) to issue a warning to developers when a deprecated operation is called in client code. Technical communication is proactive and similar to the ``@Deprecated'' mechanism in Java. The second type is \textit{non-technical communication}. Non-technical channels include API documentation/OpenAPI specification (i.e., the focus of RQ1-3), evolution-related channels (release note/changelog/updates/migration guide), and social media platforms (i.e., blog, Twitter, and API forum/community). 
We searched for the deprecation keywords in the possible communication channels, i.e., API official websites and social media accounts, to identify additional communication channels used for API deprecation. 
Thus having a twitter API account does not mean the API communicates deprecation via Twitter, unless it mentioned deprecation information using the twitter account. Note that API providers might provide email subscription service for delivering deprecation information. However, as we have no access to those email account, we could not verify if any discussion is taken there with respect to API deprecation. Thus we do not include email as a communication channel to inspect in RQ4.
 
\noindent{\bf Results.} Table~\ref{tab:rq4} represents how different communication channels are adopted for communicating deprecation in the studied 219 APIs. Note that as the 219 APIs either set the deprecated field as true or specify deprecation information in their corresponding OpenAPI specification, it is not surprising that all of them specify deprecation information in their official documentation as well. However, besides the documentation, we only find 61 APIs use a second communication channel, most of which use evolution-related communication channels for deprecation notification. Moreover, only three APIs have adopted a proactive technical communication channel: {\em Jira} uses a custom HTTP header to send out deprecation notification when a deprecated operation is called in client code; {\em Kubernetes} uses HTTP warning header (currently deprecated by HTTP~\cite{warningHeader}); {\em NBA Stats} uses error message associated with response code 404 to indicate a deprecated operation. Our results indicate that most of the communications on deprecation are in a reactive way, i.e., API documentation. While API consumers can regularly check the communication channels, we believe more proactive support is needed to keep API consumers informed of deprecation.

\begin{table}[ht]
    \centering
    \caption{Number of APIs adopting each depreaction communication channel in addition to documentation.}\label{tab:rq4}
    \begin{tabular}{p{3.5cm}p{4cm} }
    \hline
     \textbf{Communication Channel} & \hfil\textbf{Number of APIs}  \\
    \hline
API Response          & 3                                                       \\
Evolution-related~Channels   & 57 (40 release note, 9 change log,  \\
& 5 updates, 3 migration guide)  \\
Blog                  & 20                                                      \\
Forum/Community       & 6                                                       \\
Twitter               & 4     \\ 
    \hline
    \end{tabular}
    \label{tab:results}
\vspace{-0.2cm}
\end{table}
\begin{tcolorbox}
RQ4-Findings:
Only three studied APIs adopted a proactive communication channel, i.e., API consumers receive warnings when deprecation-related operations are used. In addition to documentation,  other reactive communication channels are also leveraged such as release notes and Twitter, but less common.
\end{tcolorbox}
\vspace{-0.4cm}

%% file: threats.tex
\noindent{\bf External Validity.} Our findings are based on the web APIs in APIs.guru and thus may not generalize for all web APIs. Also, the evolution analysis is based on the multiple-version specifications in APIs.guru, which may not contain all the versions of its collected APIs. However, APIs.guru is the largest directory of RESTful APIs available and is widely used in literature Future effort should curate more web API artifacts.

\noindent{\bf Internal Validity.} First, our study relies on the quality of OpenAPI specifications in APIs.guru. However since most of the specifications are official (i.e., from API providers), we consider them of high-quality. In particular, among all the web API versions we studied, only six are from unofficial sources. We manually examined the six specifications and include high-quality ones. 
Second, given the diverse deprecations (e.g., on method, request parameter), we concluded the deprecation status at operation level, e.g., if any of the operation's API elements describes the replacement, we consider this operation provides replacement message. As a result, our conclusions are upper bound of the real situation.
Third, our study focuses on the operations in web APIs, i.e., operation, request parameters, response arguments, and do not consider changes in other parts of the specifications, e.g., security and license. 
Last, both RADA and the detection of deprecation information are heuristic-based and may have inaccuracies. Our evaluation confirms both have acceptable performance.

%% file: related.tex
\noindent{\bf Studies on Web API Evolution.}
Researchers have performed studies to understand the evolution of web APIs. Prior studies~\cite{sohan2015case, li2013does} characterize the evolution patterns of web APIs.
Wang et al. characterized web API changes from 25 API versions of 11 unique RESTful APIs and tried to understand developers' reactions on the changes~\cite{wang2014developers}. They found that adding API operations is the most popular API change type and deleting API operations leads to the strongest reactions from developers on StackOverflow. 
Given the complexity in web API evolution, automated tools are proposed to support API providers in web API evolution~\cite{polak2015advanced,campinhos2017type} and to assist API consumers in managing evoving web APIs~\cite{mateos2015tool,serrano2017linked,arcuri2019restful,bae2014safewapi,wang2011apiexample,thomchick2018improving,adeleye2019fitness}. 

Different from the above work, we focus on a unique aspect in web API evolution, i.e., API deprecation, which is the standard practice in other ecosystems, yet has not been examined for its current practice in web API.

\noindent{\bf Studies on the Deprecation of Non-web API.} 
Kapur et al. found that deprecated entities are not always removed eventually while removed entities are not always deprecated. beforehand~\cite{kapur2010refactoring}. Zhou and Walker investigated API deprecation in Java frameworks and libraries and found that the deprecated-removed protocol is often not followed and removed APIs may resurrect~\cite{zhou2016api}. Sawant et al. identified 12 reasons for the deprecated features by the API producers from the analysis of four Java frameworks~\cite{sawantReasons}. Brito et al. 
found that on average, 66.7\% of the deprecated methods provided replacement messages in Java projects, the same ratio is 77.8\% in C\# projects~\cite{brito2018use}. 
Recently, Li et al. analyzed the deprecation in Android APIs and found that deprecated Android APIs are not always documented~\cite{li2018characterising}. They also reported that 78\% of the deprecated Android APIs have replacement messages. Nascimento et al. assessed API deprecation in JavaScript and found that that the use of deprecation is low but the ratio of having replacement messages is high (67\%)~\cite{nascimento2020javascript}. 
Developers' reactions to non-API deprecation are also studied. Robbes et al. conducted studies on developers' reactions to API deprecation in a small talk ecosystem~\cite{robbes2012developers}, Java projects~\cite{hora2015developers} and Java SDK~\cite{ sawant2018reaction}. They found that 61\% client systems are potentially affected by the API changes in Java SDK. Hora et al. show that 53\% of their analyzed API changes caused a reaction in only 5\% of the client applications~\cite{hora2015developers}. 

Differently in this work, we study web API deprecation and proposed RADA to detect deprecated APIs in OpenAPI specifications. Our study reveals important maintenance challenges associated with web API deprecation. It remains as future work to thoroughly investigate the negative impacts of web API deprecations on client code and study consumers' reactions to the deprecations.


%% file: conclusion.tex
The web API ecosystem has grown significantly. Breaking changes are inevitable when
web APIs are under rapid evolution.  Deprecated-removed protocol, which is the standard practice in other ecosystems (e.g., Android), should be adopted to alarm web API consumers on deprecating web APIs.

We present the first empirical study on the deprecation of web APIs. 
We propose RADA to analyze deprecated API elements in OpenAPI specifications.
Our study includes the largest web API directory {\em APIs.guru}, i.e., a total of 1,368 RESTful APIs. 
Our study reveals that the majority of the studied RESTful APIs do not follow the deprecated-removed protocol.
Our study further reveals that there are inconsistent deprecation behaviors  within one API and one organization.
Furthermore, we find that the deprecation impact is not trivial, i.e., for 32.4\% of the studied deprecation-related APIs, all of their operations are affected by deprecated API elements.
Last, we study additional knowledge and channels that API providers may communicate with API consumers. We find that even for the most frequently-provided knowledge (i.e., alternatives), only 45\% of the studied APIs suggest alternatives consistently for all deprecation-related operations. Among the additional communication channels, proactive approaches such as API response are rarely used. Most APIs utilize reactive approaches such as blogs or release notes to inform API consumers of deprecation.

Our work reveals critical maintenance challenges that API consumers may face due to the imperfect web API deprecation practice. Future research should further study the impact of API deprecation on client applications and provide automated support to facilitate and enforce the deprecation practice.


%% file: acknowledgement.tex
We acknowledge the support of the Natural Sciences and Engineering Research Council of Canada (NSERC), [funding reference number: RGPIN-2019-05071].
\clearpage